Article

# Quantum Mechanics: Statistical Balance Prompts Caution in Assessing Conceptual Implications

Brian Drummond

Independent Researcher, Edinburgh, UK; drummond.work@phonecoop.coop

**Abstract:** Throughout quantum mechanics there is statistical balance, in the collective response of an ensemble of systems to differing measurement types. Statistical balance is a core feature of quantum mechanics, underlying quantum mechanical states, and not yet explained. The concept of "statistical balance" is here explored, comparing its meaning since 2019 with its original meaning in 2001. Statistical balance now refers to a feature of contexts in which: (a) there is a prescribed probability other than 0 or 1 for the collective response of an ensemble to one measurement type; and (b) the collective response of the same ensemble to another measurement type demonstrates that no well-defined value can be attributed, for the property relevant to the original measurement type, to individual members of the ensemble. In some unexplained way, the outcomes of single runs of a measurement of the original type "balance" each other to give an overall result in line with the prescribed probability. Unexplained statistical balance prompts caution in assessing the conceptual implications of entanglement, measurement, uncertainty, and two-slit and Bell-type analyses. Physicists have a responsibility to the wider population to be conceptually precise about quantum mechanics, and to make clear that many possible conceptual implications are uncertain.

## 1. Introduction

### 1.1. Context and Overview

The concept of statistical balance is central to assessing the conceptual implications of quantum mechanics. This has been explored generally [1] (the "2019 review") and specifically in relation to violation of Bell inequalities [2] (the "2021 article"). Statistical balance pervades quantum mechanics, but most authors seem to accept that explaining this concept is outwith the scope of quantum mechanics. It is surprising, therefore, that many authors seek, or claim to have found, conceptual explanations for other aspects of quantum mechanics such as entanglement, measurement, uncertainty, and the two-slit and Bell analyses. This article highlights this apparent inconsistency.

The 2019 review and the 2021 article use only non-mathematical language, focus on conceptual analysis and give unusually extensive reference listings. The rest of this first part considers the significance of these methods, and notes to what extent they are used in this article.

The second part introduces statistical balance as a core characteristic of quantum mechanics, underlying the concept of a quantum mechanical state. In the light of this unexplained balance, the third part cautiously assesses the conceptual implications of entanglement, measurement, uncertainty, two-slit phenomena and Bell inequality violations. The final part concludes, and notes consequent possibilities and responsibilities.

While some of this article collates what has already been well-explored elsewhere, two elements of what follows represent essentially new material.

- Sections 2.1–2.3 include a much fuller account of the concept of statistical balance. This term does not appear to have been used between its original introduction in 2001 and its re-introduction in the 2019 review and the 2021 article. Informal (unpublished) responses to this reintroduction suggested that the account of statistical balance in the 2019 review and the 2021 article did not answer all the questions which it raised in readers' minds. Sections 2.1–2.3 aim to answer these questions.
- Section 4.2 refers to ideas which have been separately developed in a range of contexts, but which have not, to my knowledge, been brought together in this way in the context of quantum mechanics.

*1.2. Concepts and Language*

Conceptual analysis is an aspect of philosophy, but it is also necessary for physics [3] (p. 484) [4,5] [6] (p. 82) [7] (p. 204) [8] (Section 7) [9] (p. 29). Indeed, in dealing with the foundations of physics, it is not possible to draw any definite line between physics and philosophy [7] (pp. 7–10) [10] (p. 108) [11] [12] (p. 18): the two disciplines overlap while remaining distinct [13]. It is common for authors to dismiss the need for a conceptual "interpretation" of quantum mechanics as unnecessary, or not intrinsic to the theory. In practice, however, all quantum mechanical analysis will involve some degree of conceptual interpretation, whether or not this is not explicitly acknowledged [14] (Section 8) [15] (p. 176) [16] (p. 2) [17] (p. 5) [18]. Indeed, such analysis can provide a basis on which to develop such interpretation [19]. This article does not challenge the widely-agreed mathematical features of quantum mechanics. This article does, however, challenge some of the less-widely-agreed conceptual implications which are claimed for quantum mechanics.

Because most of these claims are made in non-mathematical language, most of this article is written in non-mathematical language. The use of non-mathematical language can be compatible with the precision necessary in physics, but this requires (a) an initial specification of how elements of language that can take different meanings are being used in any context and (b) careful and disciplined use of such language thereafter [6] (pp. 74–75). Without such specification and precision, non-mathematical language can lead to "confusing, and sometimes contradictory literature…which serves in many cases to baffle rather than enlighten" [20] (p. 3).

To meet the first of these requirements, the intended meanings of words in this article are as in the glossary in [1], as added to and amended by [2] (pp. 59–60). Three further definitions are needed for this article: the intended meaning of **concept** is a component of a structured mental representation [21] (p. 563); **conceptual analysis** is intended to mean analysis involving concepts; and the intended meaning of **event time** is the time at which particular observed events occur [22] (p. 8). All these meanings may differ from the meanings intended by other authors using the same words. Thus, for example, although the words "ensemble" and "event" can be defined in algebraic terms [23] (ch. 5) [24] (pp. 29, 32), in this article they are used in a more "physical" way [1] (p. 422) [2] (pp. 59–60).

Non-mathematical conceptual analysis in physics must be clearly supported by, and consistent with, relevant mathematical analysis. This can be demonstrated by citing sources which provide relevant mathematical analysis [7] (pp. 11, 19–20), as is done in this article. All conceptual analysis in physics also necessarily involves underlying philosophical assumptions [3] (p. 485) [7] (pp. 17–18, 204) [25] (p. 55) [26] [27] (pp. 729, 736) [28] (p. 201), and it helps when these are explicitly acknowledged [1] (p. 382) [29] (pp. 603–605). This article adopts the assumptions outlined in [1] (Sections 1.5–1.9). In particular, this article considers only nonrelativistic quantum mechanics, which treats time as independent of space [1] (p. 394).

The use of purely non-mathematical language raises the question: which non-mathematical language? Like the 2019 review and the 2021 article, this article is (a) written in English, and (b) directly reflects only literature in, or translated into, English. This limits its "epistemic diversity" [30], and contributes to systematic bias [31]. I apologise for this sad result of my current time and resource constraints.

*1.3. Citations and References*

The literature dealing with conceptual challenges arising in quantum mechanics is vast. There is no clear consensus within that literature on how to resolve these challenges. These two facts create a significant difficulty: how can authors adequately show how their particular contribution relates to the wider literature? Many themes in this article are more fully explored in the 2019 review [1] and the 2021 article [2].

The 2019 review and 2021 article are unusual in that they explicitly state that their reference listings aim to refer "to enough of the literature to be representative" [1] (p. 392) [2] (p. 48) and cite, respectively, 570 and 265 references. Even within the English language literature, these numbers represent a small fraction of the relevant material. This article does not re-cite all the references which are relevant to this article. Instead, where sections of the 2019 review and 2021 article are cited in what follows, the works cited in those sections of of the 2019 review and 2021 article should be treated as supplementing the citations here. The 190 works cited in this article are mainly those published in the last few years, and over 120 of them were cited in neither the 2019 review nor the 2021 article.

The reference listing in this article is, therefore, again unusual for an article dealing with the core concepts of quantum mechanics. The reference listing is intended for those who already have a good understanding of quantum mechanics, and at least some awareness of the wide range of approaches to its possible conceptual implications. It will allow such readers to see how my comments relate to the existing literature, and might also point them to some areas of that literature with which they might be less familiar.

The cited references are not necessarily an appropriate "further reading" list for those new to quantum mechanics, who are seeking a first impression of its conceptual implications. For such readers an alternative approach to "further reading" might be to concentrate on two or three "good textbooks". For those new to the area, which of these approaches to further reading is taken is a matter of personal preference: both approaches are potentially dangerous. Reading only the diverse, and mainly recent, journal articles cited in this article, risks attempting to build without laying a foundation. Concentrating on only a few textbooks, risks remaining unaware of persuasive challenges to the views expressed therein, or of developments subsequent to their publication. It also involves the difficult task of identifying textbooks which are "good", in the sense of dealing well with the possible conceptual implications of quantum mechanics. Identifying textbooks which are "good" in this sense is outwith the scope of this article, but the few textbooks included in the reference listing would be among the plausible candidates.

**2. Statistical Balance**

*2.1. Statistical Balance: Prescribed by Quantum Mechanics*

Quantum mechanics allows prescription ("writing in advance") of some aspects of possible future events involving physical systems [32] (Section 7). These prescriptions will usually cover a range of different, often mutually exclusive, physical contexts [1] (Section 3.2). The prescriptions are made as probability distributions, which can sometimes be applied to single events [33] (p. 9). That said, the prescriptions can *usually* only be checked by reference to a statistical sample of events [8] (p. 1678) [34] (Sections 4.2.3 and 6.4) [35] (p. 224) [36] (p. 202) [37]. The focus on prescriptions here implies no particular view of scientific theories [38], and allows for the possibility that a descriptive theory might underlie quantum mechanics.

In considering conceptual implications of quantum mechanics, therefore, this article focuses on ensembles of systems, where each ensemble is assumed to be a statistical ensemble. A statistical ensemble here is intended to mean a set of systems which can be treated as identical, such as those prepared in an identical way [39] (Section 10.1.3). The fact that the individual systems within the ensemble can be treated (for statistical purposes) as identical does not, however, necessarily imply that the systems are identical in every respect [40] (p. 2). This pragmatic focus on ensembles implies no particular view of any ontology underlying quantum mechanics, nor does it imply instrumentalism.

Core to the analysis which follows is the concept of statistical balance [1] (Section 2.2). Statistical balance refers to a feature of the many quantum mechanical contexts in which the following two facts arise: (a) there is a prescribed probability other than 0 or 1 for the collective response of an ensemble to a particular measurement type; and (b) the collective response of the same ensemble to a different measurement type demonstrates that no well-defined value can be attributed, for the property relevant to the original measurement type, to individual members of the ensemble. The combination of these two facts suggests that somehow the outcomes of single runs of a measurement of the original type must "balance" each other to give an overall result in line with the prescribed probability (which is not 0 or 1).

To illustrate this, consider two measurement types: T1, with possible outcomes O1 and O2; and T2, with possible outcomes O3 and O4 [41] (Section 8) [42] (Section 2) [43] (Section 1.1.2) [44] (Part 3).

- If, for an ensemble, the prescribed probability, that O1 is the result of a T1 measurement, is 1, then the outcomes in a sequential-in-time series of single runs of a T1 measurement will all be O1.
- Typically, for a T2 measurement on that same ensemble, the prescribed probability, that O3 is the result, is not 0 or 1. This implies that there will be differing outcomes in a sequential-in-time series of single runs of the T2 measurement.
- A T2 measurement on the ensemble effectively produces two subensembles. For one subensemble, the probability, that O3 is the result of a T2 measurement, is 1. For the other subensemble, the probability, that O4 is the result of a T2 measurement, is 1.
- If a T1 measurement is then made on each of these subensembles then, for each subensemble, the result is consistent with a prescribed probability, that O1 is the result of a T1 measurement, which is not 0 or 1. For the originally prepared ensemble, however, the prescribed probability, that O1 is the result of a T1 measurement, was 1. This implies that the the originally prepared ensemble was empirically different from the aggregate of the two subensembles resulting from the T2 measurement.
- Can a well-defined value be attributed, to the property relevant to the T2 measurement, for individual members of the originally prepared ensemble? If such values could be attributed, there is no obvious reason for the empirical difference (with respect to a T1 measurement) between the originally prepared ensemble and the aggregate of the two subensembles resulting from the T2 measurement. On this basis, it appears that no well-defined value can be attributed, to the property relevant to the T2 measurement, for individual members of the originally prepared ensemble.
- Thus, there must be some other explanation for how the outcomes of the sequential-in-time series of single runs, of a T2 measurement on the original ensemble, "balance" with each other to give an overall result in line with prescribed probability which is not 0 or 1.

### 2.2. Statistical Balance: A Specific Example

The phenomena appearing in a sequence of Stern-Gerlach arrangements, [43] (Section 1.1), reflect statistical balance, although there are (a) many unanswered questions about the concept of spin in quantum mechanics [45–48], and (b) challenges in appropriately interpreting the phenomena arising in Stern-Gerlach arrangements [49–53]. Referring to the illustration in Section 2.1 above, let T1 and T2 be two versions of a Stern-Gerlach experiment.

1. T1: systems are directed towards a vertically-aligned magnetic field, and are subsequently detected above (up, U) or below (down, D) their initial vertical level.
2. T2: systems are directed towards a horizontally-aligned magnetic field, and are subsequently detected to the left (L), or right (R), of their initial horizontal position

An ensemble is prepared so that, for a hypothetical T1 experiment, the prescribed probability of U is 1. In the actual T2 experiment, the prescribed probability of L is 0.5. Considering each of the subsequent T1 experiments in isolation, the prescribed probability of U is 0.5. Considering the T2 experiment and both of the subsequent T1 experiments together, there are 4 possible outcomes: LU, LD, RU, RD. The prescribed probabilities for each of these 4 outcomes are all 0.25. Thus, for the full sequence of experiments, the prescribed probability of [either LU or RU] is 0.5. Thus, the (aggregate) ensemble after the T2 experiment, is empirically different from the ensemble prior to the T2 experiment.

In this article, statistical balance refers to the way in which, considering the T2 experiment in isolation, the outcomes of the individual runs "balance" with each other to give an overall result in line with the prescribed probability. The subsequent T1 experiment suggests that no well-defined value can be attributed, to the property relevant to the T2 measurement, for individual members of the originally prepared ensemble. Thus, there must be some other, as yet unknown, explanation for this statistical balance.

*2.3. Statistical Balance: Pervasive but Unexplained*

This statistical balance pervades quantum mechanics. In particular, it features even in the analysis of single, simple systems (in other words, systems which have no internal structure or subsystems) [1] (Section 2.2) [41] (Section 8) [42] (Section 2) [54] (pp. 10–11). There is always a sense of balance in the statistics. In the above example, the balance is among a series of measurement event outcomes which happen at the same place but sequentially in time (where a "measurement event" is a single run of a measurement—see Section 3.2 below—and "time" here is "event time"—see Section 1.2 above). This is an instance of **two-measurement-type statistical-balance-in-time** [2] (Section 2.2). The balance can also arise among a group of measurement event outcomes which happen simultaneously but at differing places in space. Such balance might be referred to as **two-measurement-type statistical-balance-in-space**. Indeed, the balance would also arise in the analysis of a collection of measurement event outcomes which happen at different places in space, and also sequentially in time. Such balance might be referred to as **two-measurement-type statistical-balance-in-space-and-time** .

The term statistical balance was originally introduced around 20 years ago, in relation to a further aspect of the collective response of an ensemble, to differing measurement types [44] (Part 3). As noted in the example in Section 2.1 above, for each of the two subensembles resulting from the T2 measurement, the prescribed probability, that O1 is the result of a T1 measurement, is not 0 or 1. In fact these two prescribed probabilities complement each other: they add to give 1. It is a basic feature of probability that, for any given ensemble, the probabilities all the possible outcomes will add to give 1. Here, however, there is an extra feature: for the two *different* ensembles resulting from the T2 measurement, the prescribed probability for each, that O1 is the result of a T1 measurement, also add to give 1. This is well established, both mathematically and empirically, but it is conceptually surprising. The term statistical balance was introduced in 2001 to refer to this form of "double stochasticity" [44] (Part 3): the outcomes of the runs of a T1 measurement, on one subensemble resulting from the T2 measurement, "balance" with the outcomes of the runs of a T1 measurement, on the other subensemble resulting from the T2 measurement (in that they are consistent with prescribed probabilities that add to give 1). This article will not comment further on this **double-stochasticity statistical balance**.

Statistical balance arises in a group of measurement event outcomes, despite none of them being explicable in isolation from the others (for example in terms of pre-measurement underlying properties). The balance emerges in the overall result of the measurement (taking the single-run outcomes together) consistently being in line with quantum mechanical prescriptions (although it can be very challenging to unequivocally demonstrate this in practice [55,56]). The fact that the overall result is predictable is hard to explain unless, for example, a statistical description is seen as being as fundamental as a single-run description [36] (pp. 203, 255) [57].

Two-measurement-type statistical-balance-in-time has been recognized as a fundamental feature of all quantum mechanics [42] (Section 2), but most authors appear to accept, explicitly or implicitly, that the scope of quantum mechanics does not extend to explaining this pervasive balance [42] (Section 3). Relatively few authors explore possible conceptual explanations of the statistical balance. Among those who do, some propose a conservation principle operating across an ensemble, rather than for each of its members [57]; others explain the balance as a feature of independent reality, by involving concepts such as potentia(lity) and propensity [2] (p. 50) [22] (Section 4) [58] (Section 14.3) [59] (pp. 24–27), or irreversibility [36] (pp. 210, 242, 254).

Thus, quantum mechanical analyses, of even the simplest subatomic phenomena, involve a concept – statistical balance – which currently has no widely accepted explanation. This fact motivates a cautious approach to the conceptual analysis of more complex aspects of quantum mechanics. Other facts also prompt such caution [1] (Section 2.1). For example, all quantum mechanical prescriptions are for events, not for systems; so events might be, ontologically, more fundamental than systems [60–62] [63] (p. 8).

*2.4. Statistical Balance Underlies the State Concept*

In this article, the word state is intended to refer to the mathematical term which is a core element of the mathematical structure, or formalism, which is being used for quantum mechanical analysis. Some authors use "state" to mean "a material field or physical entity" [64], but this article will use the word system to refer to objects which are unquestionably physical [1] (p. 423).

Even among authors who use state to refer to a mathematical term, some explicitly suggest that the concept of state is not only mathematical, but also ontological (existing in independent reality). Other authors use language in a way that, intentionally or otherwise, suggests that the mathematical state is also ontological. Attempts to narrow the range of possibilities are ongoing [65,66], but there is little agreement in this area [67]. Thus, it remains true that "there do not yet appear to be clear grounds to accept any simple relationship between the quantum mechanical state and independent reality" [1] (p. 401) [40] (p. 6) [64] (Section 2) [68] (p. 3) [69] (Sections 20.3–20.5). One recent suggestion that might find broad support is that the state "represents something that is the reason why quantum systems behave the way they do" [70] (p. 291). This suggestion allows for the possibility that the (mathematical) state may only partly or imperfectly or indirectly represent (an ontological) something that makes quantum systems behave the way they do [71,72].

There is ongoing disagreement on whether a state necessarily relates to an ensemble, or can relate to a single system [8] (p. 1678) [73] (Section 1) [74] (Section 2). Most authors would agree, however, that the mathematical state prescribes probability distributions for collective outcomes of future measurement processes [69] (Sections 20.2 and 20.6). This article adopts the "comprehensive characterization of a quantum mechanical state" in the 2019 review [1] (p. 399). That characterization noted that "For measurement, the state prescribes, for each type, probability distributions (for outcomes of repeated measurement events of that type on an ensemble of systems) reflecting a statistical balance in collective outcomes, both within ensembles, and among ensembles for differing measurement types" [1] (p. 399). Consistent with the comment in Section 2.1 above, this characterization implies no particular view of any ontology underlying quantum mechanics, nor does it imply instrumentalism.

## 3. Entanglement, Measurement, Uncertainty, Two-Slit and Bell-Type Analyses

*3.1. Entanglement: Statistical Balance among Subsystems*

A superposition is a state formed by mathematically combining other quantum mechanical states. This does not necessarily imply physically combining systems. The mathematical form of the superposition state contains both (a) terms representing possible measurement event outcomes and (b) terms specifically representing the statistical balance between such outcomes [1] (p. 401) [60] (Section 3). The terms specifically representing the statistical balance are known as interference terms, and their presence is referred to as coherence [75,76].

If systems are to be considered together in quantum mechanics, then any separate states (representing the uncombined systems) must be replaced by a single one for the composite system [41] (Section 15). Typically, the composite system state is a superposition of composite system states, and is referred to as an entangled state [25] (p. 149). In addition to the statistical balance which characterizes all states, entangled states feature statistical balance among collective outcomes, for differing measurement types, on far-apart subsystems. This is a further instance of **two-measurement-type statistical-balance-in-space-and-time** [2] (Section 2.3).

*3.2. Measurement, But of What?*

In this article: "*measurement event* refers to the interaction of a single member, S, of an ensemble of systems, with an apparatus, A . . . *measurement* refers to a series of repeated measurement events (single runs), on members, S, of an ensemble of . . . systems . . . [39] . . . [and] the collective outcomes of the measurement events constitute the *result* of the measurement" [1] (p. 403) [77] (Section 5.3).

The result of a measurement does not necessarily indicate anything about the systems before the measurement [1] (p. 403) [34] (Section 4.6.1) [37] [77] (Section 5.3). This makes sense in the context of statistical balance: as noted above, one feature of that balance is that, very often, no well-defined value can be attributed, for the property relevant to a given measurement type, to individual members of the ensemble.

This and other features of measurement [1] (p. 403), underline the need for caution in assessing possible conceptual implications of quantum mechanics. There are, however, some areas in which insufficient analysis can result in undue hesitancy in dealing with these implications. One such area is the "measurement problem" [78] (ch. 11). The extent to which any such problem arises depends on the approach taken [34] (Section 3.1.2) [77] (Section 5.3) [79–82]. One of the main "problems" is how a unique outcome can arise from an entangled state in each measurement event.

The entangled state, as a superposition state, includes interference terms. As noted above, the presence of interference terms is referred to as coherence. The processes leading to the disappearance of such terms are therefore called decoherence [75,83,84]. Something more than decoherence is needed, however, to explain a unique outcome for any a measurement event [79] (Section 1) [85–87]. Many analyses of the measurement problem acknowledge this and go no further, but there are several approaches which go beyond decoherence to explain how a unique outcome can arise in a single run [39] (Section 11.4) [78] (ch. 11) [85] (Section 1.2) [88] (Section 12.7). Such analyses thus solve the traditional measurement "problem" but do not, however, explain the particular unique outcome that arises in any given run, nor indeed the statistical balance among these outcomes.

*3.3. Uncertainty: About Statistics or Systems?*

Among the several groups of "uncertainty relations" [89], the form and content of the Kennard-Weyl-Robertson relations [89] (Section 3.3.1), as rigorously derived from the quantum formalism, is well-established and widely-agreed. In sharp contrast, there is ongoing debate on the form and scope of other groups of relations, such as the Heisenberg noise (or error) disturbance uncertainty relations and the Heisenberg joint measurement uncertainty relations [1] (Section 4.6) [25] (Section 7.3) [34] (Section 7.10) [77] (Sections 3.1 and 4) [89] (Sections 3.3.7 and 3.4).

Conceptually, the Kennard-Weyl-Robertson relations imply the impossibility of preparing an ensemble of systems, for which the statistical spreads, in outcomes for the relevant two measurement types, have a product less than a specified lower bound [25] (Section 7.3) [34] (Section 1.7.1) [54] (p. 14) [90] (Section 6.4). These relations thus reflect specifc aspects of the statistical balance in the relevant ensemble.

There is not necessarily any "uncertainty" involved in the Kennard-Weyl-Robertson relations [77] (Sections 3.2 and 4), other than, perhaps, uncertainty over whether or not the relations involve uncertainty. Possible implications of the Kennard-Weyl-Robertson relations which might involve "uncertainty" remain unclear. For example, the relations do not necessarily imply anything about whether or not properties relevant to the measurement types might have definite values in any given member of the ensemble [25] (Section 7.3) [91] (Sections 3.2 and 3.4). It is possible that the relations do conceptually apply to single members of the ensemble, but this has not yet been formally established, and so debate on this point continues [1] (Section 4.6).

*3.4. Two-Slit Phenomena, and Underlying Assumptions*

The phenomena of the archetypal two-slit experiments are well known [34] (Section 4.5.2). In some ways these reflect another instance of two-measurement-type statistical-balance-in-time, where placing (or not placing) a detector at one of the slits represents one (or the other) type of measurement. Mathematically, there is a direct relationship between the two-slit phenomena and at least some uncertainty relations [92–94]. If there is such a relationship between two-slit phenomena and the Kennard-Weyl-Robertson uncertainty relations, this may reflect an aspect of the statistical balance underlying them both. The statistical balance reflected in the two-slit phenomena is particularly startling when the experiment is run so that the build up of the interference pattern is very slow. Perhaps for this reason some authors see the conceptual explanation of this balance as within the scope of quantum mechanics [59] (pp. 20–21).

The assumptions underlying many conceptual analyses of the phenomena are, however, open to challenge [1] (Section 5.1). For example, a detector interacting with a system does not, in itself, provide evidence of what (if any) route in physical space that system has followed [5] [59] (p. 20) [95] (p. 10) [96] (Sections II.B–II.C) [97] (p. 12).

References to "wave-particle duality", in this and related contexts, are often imprecise, inappropriate or both [97,98]. This is partly because there are several unresolved challenges in the efforts to characterize the concept of "particle"' in quantum mechanics [34] (Section 2.4.4) [98] (fn. 10, Section 6) [99–101] [102] (pp. 425, 431–446, 440).

Another misleading suggestion is that normal probability rules cannot account for two-slit interference; but all that is needed is a separate probability space for each context [1] (p. 408) [17] [103] (Section 9) [104]. This is effectively an extension, rather than an interpretation, of the normal approach to conditional probability [105] (p. 120).

Thus, the conceptual implications of the archetypal two-slit phenomena depend on what assumptions underlie their analysis [106], and none of the several possible explanations for these phenomena has yet attracted consensus [1] (Section 5.1) [107]. Analyses of the more recent variations on the two-slit experiments, and suggestions for further such variations, are also ongoing [108–110].

*3.5. Bell-Type Analyses: Few Clear Implications Emerge*

Starting from the Einstein-Podolsky-Rosen analysis [1] (Section 5.2) [111], Bell considered the conceptual implications of the quantum mechanical analysis of a widely-extended composite system entangled state [112]. In simple terms, the violation of the resulting Bell inequalities reflects the type of two-measurement-type statistical-balance-in-space-and-time discussed in Section 3.1 above [42] (Section 2) [113].

The widely-extended composite system is here reflecting the statistical balance which pervades quantum mechanics, forms of which are also seen in non-composite systems [1] (p. 409) [34] (Section 6.4) [41] (Sections 10 and 11) [114–118]. The widespread view that the scope of quantum mechanics does not extend to explaining this pervasive balance was noted in Section 2.3 above [1] (Section 2.2) [42] (Section 3). There seems, however, to be a widespread view that drawing conceptual conclusions from violation of Bell inequalities *is* within the scope of quantum mechanics. The inconsistency here is strange. Even the statistical balance arising in a straightforward context remains conceptually unexplained: two-measurement-type statistical-balance-in-time for systems which have no internal structure or subsystems (see Section 2.3 above). How, then, can we be confident about the conceptual implications of the statistical balance arising in the more complex context of violation of Bell inequalities: two-measurement-type statistical-balance-in-space-and-time for widely-extended composite systems?

There is, therefore, no obvious basis for confidence in the frequent conceptual claims that are made in this area [59] (pp. 22–24) [119]. There might be some conceptual implications of violation of Bell inequalities, but none of these are definite [2] [102] (pp. 431, 437–439) [120]. Claimed definite implications might reflect some combination of: imprecise non-mathematical language; assumptions which are, at best, open to challenge; and inadequate statistical analysis [2] [40] (Section 4) [68] (Section 7) [121] [122] (Appendix) [123] (Section 4) [124] (Section 13.5) [125–127] [128] (Section 2). For example, referring to correlations and causation in this context can be misleading [1] (Section 5.3) [129–131], and "it is a non-trivial task to distinguish between the effects of measurement uncertainties and the intrinsic statistics of the state" [132] (p. 2). A further difficulty relates to the unresolved conceptual difficulties relating to spin (Section 2.2 above), which are directly relevant to the Bell analyses [49] (Section 7).

Bell inequality violation might, at least partly, reflect (a) the contextuality of quantum mechanics (prescribed probabilities relate to a particular experimental context) [17] (Section 3) [126,133,134] and (b) data from different probability spaces being inappropriately combined [2] (p. 55) [120]. This latter possibility is strong, because Bell inequalities reflect the Boole inequality in probability theory [135] [136] (pp. 4–5). If the Boole inequality is violated then, for a system of three random variables, a joint probability distribution does not exist [2] (p. 49).

At least in principle, it might appear that some experiments to explore the conceptual implications of quantum mechanics need not involve statistical analysis. In experiments which focus on results for which there is a prescribed probability of 0 or 1, only one run of any measurement might be sufficient to demonstrate that such a prescription of the relevant theory is false. Analyses to support the same conceptual conclusions as are often drawn from Bell experiments, but which aim to avoid statistical analysis, were developed on this basis, both for composite systems with two parts [20] (ch. 6) [137], and for composite systems with more than two parts [138–140]. The latter analyses (for systems with more than two parts) are now generally identified by the acronym GHZ. As noted in Section 2.1 above, the concept of statistical balance only arises when there is a prescribed probability other than 0 or 1 for the collective response of an ensemble to a particular measurement type. Statistical balance does not, therefore, arise in GHZ contexts, which might suggest that there is more scope to draw conceptual conclusions. From the outset, however, the significance of the assumptions underlying GHZ analyses has been highlighted [139,140]. Subsequently, the need for caution in drawing conceptual implications from GHZ experiments has been emphasised [141–144], as has the intrinsic difficulty of avoiding statistical analysis in such contexts [145,146]. Attempts to develop non-statistical analyses of potential conceptual implications are ongoing [147].

The conceptual implications of the Bell-Kochen-Specker theorem are also not definite. There are several ways to understand conceptually the theorem's mathematical assumptions, and, consequently, different views on how to meet the need to give up at least one of these conflicting assumptions [1] (Section 5.7) [34] (Section 10.2.3) [148,149].

## 4. Conclusions, Possibilities and Responsibilities

*4.1. Unexplained Statistical Balance Means Prequantum Theories Remain Possible*

Throughout quantum mechanics there is statistical balance in the collective response of an ensemble of systems, to differing measurement types. This balance underlies quantum mechanical states: mathematical terms prescribing probability aspects of future events, relating to an ensemble of systems. This core feature of quantum mechanics is not yet explained. This prompts caution (a) in the conceptual analysis of entanglement, measurement and the uncertainty relations, and (b) in assessing the conceptual implications of the two-slit and Bell phenomena. To date, there are no definite implications about any physical reality underlying the statistical balance seen in an ensemble of composite systems. These conclusions leave open possibilities and underline responsibilities.

They leave open the possibility of a theory more descriptive of independent reality than is quantum mechanics. Several projects are exploring this possibility [150], including projects aiming for a theory with some form of both locality and realism [16,40] [113] (App. B) [133,151–155]. Bell-type analyses do not rule out such theories [1] (p. 421) [141,142,156], but do impose constraints on them [120,157]. Prequantum theories are also not ruled out by more recent analyses [158], and ruling them out might be impossible [157] (Section 5) [159] [160] (Section 8). For example, it has been suggested that "no-go theorems are ...best understood as go-theorems ...as a methodological starting point in theory development" [161] (p. 54). Such theory development is likely to involve rethinking some of our existing concepts [25] (p. 185), such as realism [162] (pp. 253, 255) [163] (Section 4.4). Indeed it has been suggested that "those who ...dream of ...prequantum theory will be terrified by this coming theory, by its complexity and extraordinariness; they will recall with the great pleasure the old QM-formalism, i.e., the present one, since it was so close to classical mechanics" [164] (pp. 136–137).

*4.2. Precision and Caution Needed in Discussing Quantum Mechanics' Conceptual Implications*

These conclusions also give rise to related responsibilities.

Physicists working on quantum mechanics have a responsibility to their colleagues to be conceptually precise. As noted in Section 1.2 above, conceptual analysis is a necessary part of physics, and so conceptual clarity among physicists can contribute to progress within quantum mechanics [128] (Section 1) [165,166], and its applications [167]. Physicists also have a responsibility to carefully consider the language they use in suggesting potential applications of quantum mechanics to the wider public [168,169].

Beyond the physics community, many non-physicists are interested in the conceptual aspects of quantum mechanics [170–175]. These are of direct relevance to philosophy [3] (Section 6) [176] (p. 395) [177–184], although to date their influence on philosophy has been limited [185,186]. Less directly, these conceptual aspects of quantum mechanics can be relevant to wider society [29] (p. 607) [103] [187] (pp. 770–778) [188]. Here a further responsibility arises. Physicists and philosophers writing on quantum mechanics have a responsibility to non-specialists to be conceptually precise [126] (p. 121), and to make clear that many possible conceptual implications of quantum mechanics are uncertain [160] (p. 241) [170,171,189]. Failure to do so can lead to, among other things, "the uncritical transmission of questionable aspects ...into entire interdisciplinary (sub)fields" [189] (p. 37).

That said, while cautiously acknowledging that the conceptual implications of quantum mechanics are unclear, physicists and philosophers have yet another responsibility: not to inappropriately cast doubt on clear conceptual and societal implications of other areas of physics and wider science [7] (p. 199). For example, the fact that the conceptual implications of quantum mechanics are unclear in no way detracts from the fact that we as humans are dealing with an external reality of which we are part, and in which it is important that we should act responsibly [7] (pp. 202, 268–269). Thus, in particular, philosophy of science "would do well ...to ...welcome perspectives that address the role of academic sciences and technoscience in human rights and food sovereignty issues that are, at the very least, crucially important for the well-being of more than a billion of the planet's human inhabitants" [27] (p. 735).

Overall, it is entirely appropriate and rational for physicists to hold personal or group convictions about the conceptual implications of quantum mechanics, but in presenting these, whether to fellow physicists or to a wider audience, they should "not claim for their findings a degree of rational conviction greater than what they actually have" [7] (pp. 210, 269). This calls for the doxastic form of intellectual humility which "accurately tracks ...the positive epistemic status of one's beliefs ...In other words, the intellectually humble person's perception of whether and to what degree her beliefs are justified (or true, or otherwise epistemically positive) is correct, or non-culpably incorrect" [190] (p. 650).

In the context of quantum mechanics it is appropriate to convey "epistemic humility" [13], and acknowledge that the lack of clarity on the conceptual implications of quantum mechanics "is deeply, deeply humbling, to us, even as we still experience wonder in the power of scientific inquiry" [170] (p. 18).

**Funding:** This research received no external funding

**Data Availability Statement:** Not applicable.

**Acknowledgments:** I acknowledge the work of the authors of the cited references. Their work has made mine possible. I also thank authors who have given encouragement, insight and constructive comments in response to earlier presentations of my research. These have, I hope, led to greater clarity in the present article. Finally, I thank the three anonymous reviewers: the present article has benefitted from their perceptive challenges to my initial submission.

**Conflicts of Interest:** The author declares no conflict of interest.